\documentstyle[12pt,aps,amsmath,epsfig]{revtex}
\tightenlines
\begin{document}

\title{Electron transfer in the nonadiabatic regime: Crossover from 
quantum-mechanical to classical behaviour}
\author{
  Gunther Lang, $^{1}$   Elisabetta Paladino,$^{1,2}$ 
  and Ulrich Weiss $^{1}$    }
 \address{ 
 $^{1}$ Institut f\"{u}r Theoretische Physik, Universit\"{a}t
           Stuttgart, 70550 Stuttgart, Germany\\
  $^{2}$  Dip. di Metodologie Fisiche e Chimiche per l'Ingegneria,
         Universit\`a di Catania, 
        Viale~A.~Doria~6, 95125 Catania, Italy}
\address{\rm email: gunther@theo2.physik.uni-stuttgart.de, 
elisa@meso.if.ing.unict.it, weiss@theo2.physik.uni-stuttgart.de}
\date{Date: \today}
\maketitle

\begin{abstract}
We study nonadiabatic electron transfer within the biased spin-boson model. 
We calculate the incoherent transfer rate in analytic form at all temperatures
for a  power law form of the spectral density of the solvent coupling. In the
Ohmic case, we  present the exact low temperature corrections to the zero
temperature rate for arbitrarily large bias energies between the two redox
sites.  Both for Ohmic and non-Ohmic coupling, we give the rate in the entire
regime extending from  
zero temperature, where the rate depends significantly on the detailed
spectral behaviour, via the crossover region, up to the classical
regime.  For low temperatures, the rate shows characteristic quantum features, 
in particular the shift of the rate maximum to a bias value below the
reorganization energy, and the
asymmetry of the rate around the maximum. We study in detail the gradual
extinction of the quantum features as temperature is increased.

\end{abstract}

\narrowtext

\section{Introduction}

Electron transfer plays an important role in many chemical and biological
reaction processes \cite{marcus1,Ulstrup,Fain,Benderskii,Devault}. 
A prime example is the ultrafast primary electron transfer step in
photosynthetic reaction centers, which is 
responsible for the high efficiency of the photosynthetic mechanism. 
The theory of electron transfer (ET) in chemical reactions goes back to 
Marcus \cite{marcus2} and Levich \cite{levich}. They were among
the first who recognized the importance of the solvent environment.
Since the donor and acceptor states
are strongly solvated, the transfer of the electron entails a
reorganization of the environment. This results in a free energy barrier
separating reactants and products.
The transition of the electron from the reactant to the product state
can only take place if favorable bath fluctuations bring reactant and
product states into resonance.  

Electron transfer is conveniently discussed by considering the
diabatic potential energy surfaces (PES) of each electronic state. In
the classical limit, the transfer rate is determined by two 
factors. First, a Boltzmann factor with the activation energy $E_a$
required for bath fluctuations to the crossing region of the two
diabatic PES. 
Second, an attempt frequency prefactor describing the probability for
tunnelling once the levels are in resonance. The latter factor is mainly
determined by the overlap between the electronic wave functions localized
on different redox sites. For large intersite coupling, the reaction is
adiabatic, whereas for weak intersite coupling, the reaction is
nonadiabatic. 
In this paper, we restrict ourselves to  nonadiabatic 
electron transfer, where we have fast thermal fluctuations to the vicinity
of the crossing region followed by a fast transversal of this region.
Then the intersite coupling $\Delta$ can be treated in perturbation
theory, which leads in lowest order to the Golden Rule formula for the 
rate, $k^{+} \propto \Delta^2 \exp(- E_a/ k_{\rm B} T)$. 

As the temperature is lowered to the quantum regime, nuclear tunnelling through
the free energy barrier, which crucially depends on the detailed form
of the spectral density, becomes effective and competes with
thermal activation. We concentrate on spectral
densities with a power law form at low frequencies, which has proven
to be a useful characterization in many chemical and condensed matter
applications. 
The appearance of quantum features in the ET rate with decreasing temperature
has been repeatedly studied in the last decades (see Refs.
\cite{hanggi,Siders,Bader,Wolynes,Garg} and references
therein). However, a unified analytical 
treatment of the nonadiabatic case in the entire temperature regime
is still missing. 

The purpose of the present work is twofold. First, we give a detailed study of
the quantum signatures for different spectral densities. The most
characteristic quantum features are the shift of the rate maximum to a
bias value  below the classical reorganization energy, and the
asymmetry of the rate around the maximum. Second, we discuss 
quantitatively the gradual extinction of  the quantum effects as the
temperature is tuned from the quantum to the classical regime. 
In Section II, we introduce the model and sketch the
properties of the classical Marcus rate. In Section III, we 
review the important Ohmic case, for which the deep quantum regime can be
described in analytic form.  Special attention is
paid to the situation of a large bias energy between the
donor and acceptor state for nonzero temperature.
To our knowledge, this case has not been studied systematically until
now, except for some numerical results in Ref. \cite{TangLin}. 
In the remainder, we  consider
the rate in steepest descent ranging from the sub-Ohmic to the super-Ohmic
case.  General features  are discussed in Section IV. In Section V, we study
the rate at $T=0$ and discuss the differences between the Ohmic and the
non-Ohmic case. The two remaining sections are
devoted to an analytic study of the crossover from quantum to 
classical behaviour.  In Section 
VI, we concentrate on the case of small bias energies between the two redox
sites. Finally, in Section VII, we consider the interesting  strong bias
regime, in which quantum features are most pronounced. We present a
uniform description throughout the whole temperature range.

\section{The Model}

As implied by the success of classical Marcus theory, the electron can be
described in terms of two discrete states at all temperatures of interest.
Modelling the solvent by a linearly responding heat bath,
the proper Hamiltonian is the spin boson Hamiltonian 
\cite{Chandler,leggett,book}
\begin{equation}\label{et1}
H= -\frac{\hbar}{2}(\Delta\sigma_x-\epsilon\sigma_z) +
\sum_{\alpha}\Big[ 
\frac{p_{\alpha}^2}{2m_{\alpha}}+\frac{1}{2}m_{\alpha}\omega_{\alpha}^2
x_{\alpha}^2\Big] - \mu {\cal E}\sigma_z^{} \;.  
\end{equation}
Here, $\sigma_z$ and $\sigma_x$ are  Pauli spin
matrices. Throughout, we put $\epsilon>0$. 
The donor and acceptor states are the eigenstates of $\sigma_z$ with
eigenvalues $+1$ and $-1$, respectively, and they differ by a bias energy
$\hbar\epsilon$. Transitions between these localized states are
induced by the intersite or electronic coupling $\hbar\Delta$.
The  spin operator $\sigma_z$ is coupled linearly to the
collective bath mode
$\mu{\mathcal{E}}=(q_0/2)\sum_{\alpha}c_{\alpha}x_{\alpha}$, where
$\mathcal{E}$  can be considered as a fluctuating dynamical
polarization which vanishes on average, and $2\mu$ is the
difference in the dipole moments of the two electronic states.
The effects of the solvent are fully described by the spectral density
$G(\omega)=(q_0^2/2\hbar)\sum_{\alpha}(c_{\alpha}^2/m_{\alpha}
\omega_{\alpha})\delta(\omega-\omega_{\alpha})$. 
Outer-sphere ET reactions are controlled by fluctuations
in the solvent polarization which cover a wide frequency range.
To model the solvent environment, we choose a smooth spectral density
with a power law form at low frequencies and an exponential 
cutoff at high frequencies, 
\begin{equation}
\label{spectral}
 G (\omega) = 2\delta_s \omega_{\rm ph}^{1-s}
\omega^s_{} \,e^{-\omega/\omega_{\rm c}}\;.
\end{equation}
Here $\omega_{\rm c}$ is a cutoff frequency for the bath modes and $\delta_s$
is a dimensionless coupling parameter. For convenience, we discriminate the
reference frequency $\omega_{\rm  ph}$  from the cutoff frequency
$\omega_{\rm c}$. Ohmic or frequency-independent  damping is 
described by $s=1$, whereas the regimes $s>1$ and  $s<1$ are  referred
to as super-Ohmic and  sub-Ohmic damping. 

\begin{figure}[t]
\setlength{\unitlength}{1cm}
\vspace{-5mm}
\centerline{
\begin{picture}(9,5)
  \put(0,0.5){\epsfig{file=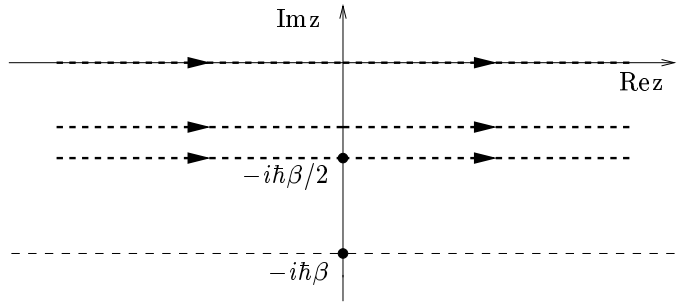,width=9cm,angle=0}}
  \end{picture}}
\caption{\label{contour}The complex time plain $z$ with three possible 
contours $\mathcal{C}_{\tau}$.}
\end{figure}

For sufficiently high temperatures and/or strong damping, the system
exhibits overdamped exponential relaxation towards the equilibrium
state. Hence the dynamics in this regime is completely described by
transition rates which obey the principle of detailed balance. 
In the nonadiabatic regime  $\Delta\ll \omega_{\rm c}$, the
incoherent tunnelling rate from the donor to the acceptor 
state (forward rate) is given by the Golden Rule
expression~\cite{leggett},
\begin{equation}
\label{rate}
k^{+}(\epsilon) = \frac{\Delta^2}{4} 
               \int_{- \infty}^{\infty} dz
               \;  e^{i \epsilon z - Q(z)}\;.
\end{equation}
In Eq.~(\ref{rate}), the system-environment coupling is considered in all
orders, whereas the electronic coupling is treated in second order.
The quantity $Q(z)$ is the twice integrated bath correlation function, 
$d^2Q(z)/dz^2=(2\mu/\hbar)^2\langle{\mathcal
E}(z){\mathcal{E}}(0)\rangle$,
and is given by \cite{leggett,book}
\begin{equation} \label{Qgen}
Q(z) =\int_0^\infty\! d\omega\,\frac{G(\omega)}{\omega^2}
\bigg\{ \coth\bigg(\frac{\hbar\beta\omega}{2}\bigg)\Big(1-\cos(\omega
z)\Big)
+ i\sin(\omega z) \bigg\}\;, 
\end{equation}
where $\beta=1/k_{\rm B}T$. The function $Q(z)$ is an analytic function of the
complex time $z$ in the strip $0 \ge \,{\rm Im}\,z > -\hbar\beta$ and
obeys the symmetry relation $Q(-z -i\hbar\beta) = Q(z)$.  
Therefore, we can arbitrarily distort the integration contour in
Eq.~(\ref{rate}) inside this strip.  
We may choose, e.g., the contour ${\mathcal{C}}_{\tau}$ with constant negative
imaginary part $0\le\tau<\hbar\beta$, as shown in Fig.~\ref{contour}.  

If we had derived 
the rate formula (\ref{rate}) with the alternative Im $\!F$-method instead of
the Golden Rule approach, we would have directly obtained the
contour $\mathcal{C}_{\bar\tau}$ where  $\bar\tau$ is the width of the bounce
(kink anti-kink pair) in imaginary time.  
The equivalence of these seemingly
different approaches follows from the analytic
properties of $Q(z)$. Putting 
$\tau=\hbar\beta/2$ and $z=t-i\hbar\beta/2$, the rate (\ref{rate}) is
equivalently  expressed as \cite{grabert}
\begin{align}\label{rateX}
k^+_{}(\epsilon) &= \frac{\Delta^2_{}}{4}\,e^{\hbar\beta\epsilon/2}_{}
\int_{-\infty}^\infty \! dt\, e^{i\epsilon t}_{}\,e^{-X(t)}_{} \;=\;
\frac{\Delta^2_{}}{2}\, e^{\hbar\beta\epsilon/2}_{}\,
\int_0^\infty \! dt \, \cos(\epsilon t)\, e^{-X(t)}_{} \;.
\end{align}
The function $X(t)$ is real and symmetric in 
$t$,
\begin{align}
\label{Xgen}
X(t)&\equiv Q(z=t-i\hbar\beta/2)=
\int_0^\infty \! d\omega \,
\frac{G(\omega)}{\omega^2_{}}\left[ \coth({\textstyle\frac{1}{2}}
\hbar\beta\omega) - \frac{\cos(\omega
t)}{\sinh(\frac{1}{2}\hbar\beta\omega)}
\right] \; .
\end{align}
In the representation (\ref{rateX}), the detailed balance relation between
the forward rate $k^+(\epsilon)$  and the backward rate 
$k^-(\epsilon)=k^+(-\epsilon)$ is directly visible,
\begin{equation}
k^+(\epsilon)=e^{\hbar\beta\epsilon}\,k^-(\epsilon)\;.
\end{equation}
With the form (\ref{spectral}), 
the integrations in Eqs.~(\ref{Qgen}) and (\ref{Xgen}) can be performed
exactly, yielding
 \begin{align}\nonumber
 Q(z) &= 2\delta_s\Gamma(s-1)
\bigg(\frac{\omega_{\rm c}}{\omega_{\rm ph}}\bigg)^{s-1}\Bigg\{ 
\Big(1 -(1+ i\omega_{\rm c} z)^{1-s}\Big) 
+  2 (\hbar\beta\omega_{\rm c})^{1-s}_{} \zeta(s-1,1 + \kappa) \\
\label{Q}
&\quad- (\hbar\beta\omega_{\rm c})^{1-s}_{}\bigg[\,
\zeta\Big(s-1,1+\kappa + i\,\frac{z}{\hbar\beta}\Big) +
\zeta\Big(s-1,1+\kappa - i\,\frac{z}{\hbar\beta}\Big) \,\bigg] \Bigg\}
\;,\\
X(t) &= 2\delta_s\Gamma(s-1) \label{X}\notag
\bigg(\frac{\omega_{\rm c}}{\omega_{\rm ph}}\bigg)^{s-1}\Bigg\{ 
 1 \; + \; 2 (\hbar\beta\omega_{\rm c})^{1-s}_{} \zeta(s-1,1 + \kappa) \\
&\quad- (\hbar\beta\omega_{\rm c})^{1-s}_{}\bigg[\,
\zeta\Big(s-1,\frac{1}{2}+\kappa + i\,\frac{t}{\hbar\beta}\Big) +
\zeta\Big(s-1,\frac{1}{2}+\kappa - i\,\frac{t}{\hbar\beta}\Big) 
\,\bigg] \Bigg\} \;.
\end{align}
Here, $\zeta(y,q)$ is Riemann's generalized zeta function \cite{Grad},
$\Gamma(y)$ is Euler's gamma function, and 
$\kappa =  1/\hbar\beta \omega_{\rm c}$. For integer $s$, Eqs.~(\ref{Q}) and
(\ref{X}) can be expressed in terms of polygamma functions. We shall refer to
the case $\kappa\ll 1$ as the quantum limit, whereas the case $\kappa \gg 1$
describes the classical limit in which $k_{\rm B}T$ is the largest energy 
scale of the problem. Note that for a high cutoff $\omega_{\rm c}$, the
quantum regime may extend up to quite high temperatures. Oppositely,
for a very low cutoff, the classical regime reaches down to low
temperatures. 

In the classical limit $\kappa \gg 1$, the integral in Eq.~(\ref{rate}) is
dominated by the short-time regime, in which we can substitute the classical
form  $Q_{\rm cl}(z)$ for Eq.~(\ref{Qgen}),
\begin{equation}\label{Qclass}
Q_{\rm cl}(z)=i\Lambda_{\rm cl}z+ (\Lambda_{\rm cl}/\hbar\beta)z^2\;.
\end{equation}
Here, $\hbar\Lambda_{\rm cl}$ is the classical reorganization energy of the
solvent, where
\begin{equation}\label{reorg}
\Lambda_{\rm cl}=\int_0^\infty d\omega\,\frac{G(\omega)}{\omega}
=2\delta_s\Gamma(s)(\omega_{\rm c}/\omega_{\rm ph})^{s-1}\omega_{\rm c}\;.
\end{equation}
The second equality holds for the spectral density (\ref{spectral}).
With Eq.~(\ref{Qclass}), the calculation of the rate (\ref{rate}) 
reduces to a Gaussian integration, giving the classical Marcus rate in 
the nonadiabatic limit
\begin{align}\label{marc}
 k_{\rm cl}^+(T,\epsilon)&=\frac{\Delta^2}{4}\sqrt{\frac{\pi\hbar\beta}
       {\Lambda_{\rm cl}}}    \exp\bigg\{-\frac{\hbar\beta}{4\Lambda_{\rm
cl}}
    (\epsilon-\Lambda_{\rm cl})^2\bigg\}\;.
\end{align}
The influence of every bath is thus characterized by 
only one parameter, the classical reorganization energy, and the rate is
insensitive to details of the spectral density. If the bias equals the
reorganization frequency, $\epsilon=\epsilon^\ast=\Lambda_{\rm cl}$,
the transfer is activationless, and the rate is maximal for fixed $T$.
Furthermore, the rate is symmetric around the maximum.

As the temperature is lowered to the quantum regime, nuclear tunnelling
through the free energy barrier
becomes effective. The quantum rate then shows some characteristic
features which sensitively depend on the detailed form of the spectral
density $G(\omega)$.  
In the next section, we review the Ohmic case $s=1$, in which the
integral in  Eq.~(\ref{rate}) or (\ref{rateX}) can be calculated
in analytic form  in the quantum regime. For
non-Ohmic spectral densities,  this is no longer possible, and we have to
employ the method of steepest-descent.

\section{The Ohmic rate in the quantum regime}
\label{cOhmic}

In the Ohmic case $s=1$, the spectral density (\ref{spectral}) takes the 
form $G(\omega)=2K\,\omega\,e^{-\omega/\omega_{\rm c}}$, where the
dimensionless  coupling strength $\delta_1=K$ is often called Kondo parameter.
By carefully performing the limit $s\to 1$ of the expressions (\ref{Q})
and (\ref{X}), we obtain
\begin{align}
\label{QO}
Q(z) &= 2K\,\ln(1+ i \omega_{\rm c} z)+
2K\, \ln\left( \frac{\hbar\beta\omega_{\rm c}\,\Gamma^2_{}(1+\kappa)}{
\Gamma(1+\kappa  +i z/\hbar\beta) \Gamma (1+\kappa 
-iz/\hbar\beta)} \right)\;,  \\ 
X(t) &= 2K\,\ln\left( \frac{\hbar\beta\omega_{\rm c}\,\Gamma^2_{}(1+\kappa)}{
\Gamma(\frac{1}{2}+ \kappa  +i t/\hbar\beta)\Gamma (\frac{1}{2}+\kappa 
-it/\hbar\beta)} \right)  \;.\label{XO}
\end{align}
In the quantum regime $\hbar\beta\omega_{\rm c} \gg 1$, the second term in 
Eq.~(\ref{QO}) is simplified and we get
\begin{align}
\label{QlT}
Q(z) &= 2K\,\ln(1+ i \omega_{\rm c}z) +   2K\,\ln \left ( 
      \frac{\sinh (\pi z /\hbar \beta)}{\pi z /\hbar \beta}\right) \;.
\end{align}
The first term represents the correct $T=0$ limit without any 
approximation. In the often employed scaling limit, the first term in
Eq.~(\ref{QlT}) is approximated by $2K\ln(i \omega_{\rm c}z)$.  The scaling
form for $Q(z)$ is only justified if the contribution of the core 
region  $|z| \leq 1/ \omega_{\rm c}$ to the rate integral (\ref{rate}) is
negligible. This is the case only if the bias is much smaller than the 
cutoff frequency, $\epsilon\ll\omega_{\rm c}$.
The equivalent scaling form of $X(t)$  reads  
\begin{align}
X(t) &= 2K \ln\left[\frac{\hbar\beta\omega_{\rm c}}{\pi}\,
\cosh\bigg(\frac{\pi t}{\hbar\beta}\bigg)\right] \; .\label{Xs}
\end{align}
With Eq.~(\ref{Xs}), the rate (\ref{rateX}) is calculated in analytic form as 
\begin{equation}
\label{ROT}
k^{+}(T, \epsilon) = \frac{\Delta^2}{4\omega_{\rm c}} 
                   \left(\frac{\hbar\beta\omega_{\rm c}}{2\pi}\right)^{1-2K}
          \frac{|\Gamma(K+i\hbar\beta\epsilon/2\pi)|^2}{\Gamma(2K)}
                    \;e^{\hbar\beta\epsilon/2}\; ,
\end{equation}
which is valid for $k_{\rm B}T, \,\hbar\epsilon\ll
\hbar\omega_{\rm c}$. The behaviour of this already well-known expression 
is discussed in Refs. \cite{leggett,book}. We remark that the form
(\ref{ROT}) was originally  obtained from the 
representation (\ref{rate}) using the scaling form  of $Q(z)$
\cite{ohmic1,ohmic3,ohmic2}. 
In this representation, however, the rate integral (\ref{rate}) seems to exist
only for  $K\le \frac{1}{2}$, since the integrand is
singular at the origin.   Actually, this singularity is spurious due to the
above analytic properties of $Q(z)$, and therefore Eq.~(\ref{ROT}) is valid
for all $K\ge 0$. In the representation
(\ref{rateX}) with Eq.~(\ref{Xs}), the singularity is far from the integration
contour for all $K$.

So far in the literature, little attention has been paid to the
case where the bias is of the order of $\omega_{\rm c}$ or larger, and
the bath is still quantum mechanical, $\hbar\beta\omega_{\rm c} \gg1$
\cite{TangLin}. If the bias is of the same order of magnitude as
$\omega_{\rm c}$, most of the contribution to the integral
(\ref{rate}) comes from the 
core region $|z| \leq 1/ \omega_{\rm c}$. Then the scaling form of $Q(z)$ is
inappropriate  and we have to use the expression 
(\ref{QlT}). With this form, however, it is not possible to
express the  integral in (\ref{rate}) in analytic form for arbitrary
temperatures. 
At zero-temperature, where $Q(z) = 2K \ln(1 + i \omega_{\rm c} z)$, the
integration in (\ref{rate}) can be performed exactly, 
yielding \cite{ohmic2,EMW}
\begin{equation}
\label{ROzero}
k^{+}(T=0, \epsilon) = \frac{\pi\Delta^2}{2\omega_{\rm c}}
\frac{1}{\Gamma(2K)}   \left(\frac{\epsilon}{\omega_{\rm c}}\right) ^{2K-1} 
                      e^{-\epsilon/\omega_{\rm c}}.
\end{equation}
The core region $|z|\leq 1/ \omega_{\rm c}$ gives rise to  the
factor $e^{-\epsilon/\omega_{\rm c}}$. 
For $\epsilon\ll\omega_{\rm c}$, this factor is unity and we
recover the $T=0$ limit of Eq.~(\ref{ROT}). For a larger bias, the
factor $e^{-\epsilon/\omega_{\rm c}}$ represents an essential
modification of the $T=0$ limit of the scaling result (\ref{ROT}).
For $K>1/2$, the rate has a maximum when the bias equals the value
$\epsilon^*= (2K-1) \omega_{\rm c}= \Lambda_{\rm cl} - \omega_{\rm c}$. In the
second form, we have used the  reorganization frequency  (\ref{reorg}) in the
Ohmic case, $\Lambda_{\rm cl}=2K\omega_{\rm c}$. Thus the position of
the rate maximum is at a lower bias value compared to the classical
regime, where  $\epsilon^*=\Lambda_{\rm cl}$.
The shift of the maximum is due to quantum mechanical nuclear
tunnelling through the free energy barrier, and the relative shift 
$\omega_{\rm c}/\Lambda_{\rm cl}=1/2K$ is small for large coupling
strength $K$.  
For $K\le \frac{1}{2}$, no rate maximum exists. However, depending on
the magnitude of the bias, electron tunnelling might be coherent and
thus a rate description may be inappropriate in this coupling regime.

At finite temperature, the contribution of the core region
to the integral (\ref{rate}) can not be given in analytic
form. However, the exact low temperature expansion for arbitrary bias
can be found. We proceed by writing Eq.~(\ref{QlT}) as
$Q(z)=2K\,\ln(1+ i \omega_{\rm c}z)+\delta Q(z)$. In the regime 
$K\,k_{\rm B} T \ll \hbar \epsilon$,  we may expand $\delta Q(z)$ 
into a power series of $(z/\hbar\beta)^{2}$,
\begin{equation} 
\label{Qlo}
\delta Q(z) = K\frac{\pi^2}{3} 
      \bigg( \frac{z}{\hbar \beta} \bigg)^2 +{\mathcal O} \bigg[\bigg(
        \frac{z}{\hbar \beta} \bigg)^4\bigg]\;.
\end{equation}
Substituting the series (\ref{Qlo}) into the integral (\ref{rate}), we 
see that every term of the resulting series for  $\exp[{-\delta Q(z)}]$
can be generated by suitable
differentiations with respect to the bias. Thus, in principle, the full low
temperature expansion can be constructed upon using Eq.~({\ref{ROzero})}. In
leading order, we obtain the exponential correction
\begin{align}
k^{+}(T, \epsilon) &= \exp\left[K \frac{\pi^2}{3} 
                       \left( \frac{1}{\hbar \beta} \right)^2 
                    \frac{\partial^2} {\partial \epsilon^2} 
                       \right]k^{+}(T=0, \epsilon)\;\\
             &=  k^{+}(T=0, \epsilon)
                \exp \left\{\frac{\pi^2}{3} \frac{K}{(\hbar \beta)^2}
            \left [ - \frac{2K-1}{\epsilon^2} + 
              \left ( \frac{1}{\omega_{\rm c}} - \frac{2K-1}{\epsilon}
\right)^2     \right ]   \right\}\;,     \label{ratelo}
\end{align} 
which  varies as $\exp({aT^2})$. Depending on the parameters, the sign of $a$
can be positive or negative. 
The contribution $1/ \omega_{\rm c}$ in the round bracket comes from the
core region. This term is negligible when the bias is 
considerably below $(2K-1)\omega_{\rm c}$, and in this regime
Eq.~(\ref{ratelo}) coincides with the low temperature expansion of the rate
(\ref{ROT}). For a stronger bias, however, this term is of crucial importance.
The maximum of the rate (\ref{ratelo}) for $K>\frac{1}{2}$ is at
$\epsilon=\epsilon^{*}$ where 
\begin{equation} \label{eps*O}
\epsilon^* =(2K-1) \omega_{\rm c}+\frac{1}{3}\frac{2K\,\omega_{\rm c}}{2K-1} 
           \left (\frac{\pi k_{\rm B}T}{\hbar\omega_{\rm c}} \right )^2\;.    
\end{equation} 
With the onset of thermal activation at finite $T$, 
 the rate maximum is shifted to a higher bias.

Let us now expand the rate expression (\ref{ratelo}) around the maximum,
Eq.~(\ref{eps*O}).  In contrast to the classical formula (\ref{marc}), we
now have an asymmetric distribution 
\begin{align}\label{expand}
k^+(T,\epsilon)&=k^+(T,\epsilon^*)\exp
      \bigg\{-\frac{(\epsilon-\epsilon^*)^2}{2W^2\omega_{\rm c}^2}
      \bigg(1-A\frac{(\epsilon-\epsilon^*)}{\omega_{\rm c}}+{\mathcal O}
      \bigg[\frac{(\epsilon-\epsilon^*)^{2}}{\omega_{\rm c}^{2}}\bigg]\bigg)
\bigg\}\;.  
\end{align}
The dimensionless width  $W$ and  the asymmetry parameter $A$ are given by
\begin{align}\label{widthO}
W^2&= (2K-1) \bigg\{1+
                \frac{2K}{3} \frac{2K-2}{(2K-1)^2}
              \bigg(\frac{\pi k_{\rm B}T}{\hbar\omega_{\rm c}}\bigg )^2
                  \bigg\}\;,\\
A &= \frac{2}{3(2K-1)}\bigg\{1-\frac{4K}{3}\frac{2K-2}{(2K-1)^2}
              \bigg(\frac{\pi k_{\rm B}T}{\hbar \omega_{\rm c}}\bigg )^2
                 \bigg\}\;.\label{asyO}
\end{align}
For $K\gg 1$, Eq.~(\ref{expand}) is accurate over a wide bias range. 
Since $A>0$ for $K>\frac{1}{2}$, the wing of the rate above the maximum is
enhanced compared to the wing  below the maximum. This is due to the smaller
width of the free energy barrier in the inverted regime $\epsilon>\epsilon^*$.
At $T=0$, we have $W^2=(2K-1)$ and $A=2/[3(2K-1)]$.
The width grows with increasing damping strength $K$, whereas
the asymmetry decreases. 
For finite temperatures, thermal activation over the free energy
barrier becomes possible. For $K>1$, this leads to a widening of the
distribution and a reduction of the asymmetry. 
The quantum features are reduced and the system behaves more
``classically'' with increasing temperature. 
In the regime $K<1$, the behaviour is opposite: the width decreases and 
the asymmetry grows for finite temperatures. However, for $K<1$ the
representation (\ref{expand}) is only accurate in a small interval around
$\epsilon^*$.

So far we have studied the Ohmic case in the quantum regime
$\kappa\ll 1$. It is now interesting to see whether 
the quantum characteristics change qualitatively for a non-Ohmic spectral
density.

\section{Evaluation of the rate in steepest descent}

For a general spectral density, it is not possible to
calculate the rate (\ref{rate}) or (\ref{rateX}) in analytic form.
To evaluate the integral asymptotically, we choose the contour
${\mathcal C}_{\bar \tau}$ which passes through the saddle point
$z_s=-i\bar\tau$  
in the direction of steepest descent (cf. the discussion in Section II
and Fig.~\ref{contour}).
Putting
\begin{equation}
\label{F}
F(z) = Q(z) - i \epsilon z\;,
\end{equation}
the saddle point $z_s$ is determined by the transcendental equation
\begin{equation}
\label{transc}
F'(z_s) = 0 \, ,
\end{equation} 
where the prime denotes differentiation with respect to the
argument. 
Now, expanding $F(z)$ around $z_s$ up to second order,
\begin{equation}\label{expa}
F(z)=F(z_s)+{\textstyle\frac{1}{2}}F''(z_s)(z-z_s)^2+{\mathcal
O}[(z-z_s)^4],
\end{equation}
we arrive at the steepest descent expression for the rate
\cite{Ulstrup,Wolynes},
\begin{equation}
\label{SDR}
k^{+}(T, \epsilon) = \frac{\Delta^2}{4} 
{\left( \frac{2\pi}{ F''(z_s)} \right )}^{1/2}
\exp[- F(z_s)] \; .
\end{equation}
The quantities $F(z_s)$ and $F''(z_s)$ are positive real. Here,
$\beta^{-1}F(z_s)$ is the ``quantum activation free energy'' of the
donor state, and $F''(z_s)$ introduces quantum effects in the
pre-exponential factor. As we shall find, the factor $F''(z_s)$ is
responsible for  the shift of the rate maximum to a bias
value $\epsilon^*$ below $\Lambda_{\rm cl}$. 
The leading correction to the formula (\ref{SDR}) comes from the fourth 
order contribution in Eq.~(\ref{expa}). This contribution is small
when the condition 
\begin{equation}\label{validity}
Q^{\rm  IV}(z_s)/8[Q''(z_s)]^2\ll 1
\end{equation}
is fulfilled. Physically, this is  the regime of multi-phonon emission 
processes.
 
In the absence of a bias, the transcendental equation (\ref{transc}) reduces 
to $Q'(z_s)=0$. With the general form (\ref{Qgen}), the stationary point
is then given by $z_s = -i \hbar \beta / 2$. 
As the bias is increased, the stationary point moves from 
$z_s = -i \hbar \beta / 2$ towards $z_s = 0$, and in general one has
to solve the transcendental equation (\ref{transc}).
The stationary point finally reaches the origin if the bias equals the 
classical reorganization frequency, $\epsilon=\Lambda_{\rm
cl}=-iQ'(0)$.

In the classical limit $\hbar\beta\omega_{\rm c}\ll 1$, the
stationary point is always near to the origin. Then the
expansion (\ref{Qclass}) around $z=0$ is justified for any bias.

\section{Zero temperature rate}

In this section, we discuss the modifications of the quantum rate features
when the spectral density is changed from Ohmic to sub-Ohmic and super-Ohmic.

For $T=0$, the pair interaction (\ref{Q}) reduces to the form
\begin{equation}\label{subo1}
Q(z) \;=\;  2\delta_s\Gamma(s-1)(\omega_{\rm c}
/\omega_{\rm ph})^{s-1}
[\, 1 - (1+i\omega_{\rm c}^{}z)^{1-s} \,]\;.
\end{equation}   
Substituting Eq.~(\ref{subo1}) into Eq.~(\ref{F}), the 
saddle point is found in analytic form, $z_s= -i [\, (\Lambda_{\rm
cl}/\epsilon)^{1/s} -1\,]/\omega_{\rm c}$. Then the  rate is obtained
from Eq.~(\ref{SDR}) as 
\begin{equation}\label{subo3}
k^{+}(\epsilon)=\; \frac{\Delta^2}{4} 
        {\sqrt\frac{2\pi}{ s \Lambda_{\rm cl}\omega_{\rm c}}}
 \left(\frac{\Lambda_{\rm cl}}{\epsilon}\right)^{\frac{1+s}{2s}}\!
\exp\bigg\{-\frac{\epsilon}{\omega_{\rm c}} + \frac{1}{s-1}\,
\frac{\Lambda_{\rm cl}}{\omega_{\rm c}}\bigg[\,s\bigg(\frac{\epsilon}{
\Lambda_{\rm cl}}\bigg)^{\frac{s-1}{s}}\! - 1\,\bigg]\bigg\}  \; .
\end{equation}
The condition (\ref{validity}) for the validity of the form 
(\ref{subo3})  reads
\begin{equation}\label{subo2}
[\,2\delta_s\Gamma(s)\,]^{1/s}(\epsilon/\omega_{\rm
ph})^{(s-1)/s}
= (\epsilon/\Lambda_{\rm cl})^{(s-1)/s}\Lambda_{\rm
cl}/\omega_{\rm c} 
\gg  (s+2)(s+1)/8s \;.
\end{equation}
This roughly corresponds to the regime 
$\epsilon \leq \omega_{\rm ph}$ in the sub-Ohmic case,
$\epsilon \geq \omega_{\rm ph}$ in the super-Ohmic case,
and to $\delta_1^{}\equiv K \gg1$ in the Ohmic case. It can  easily be
seen that the  limit $s\to 1$ in Eq.~(\ref{subo3}) reproduces the 
Ohmic rate (\ref{ROzero}) for $K\gg 1$.
The rate (\ref{subo3}) has a maximum when the bias takes the value
$\epsilon^*$, which is determined by the transcendental equation
\begin{equation}\label{emax1}
(\epsilon^*/\Lambda_{\rm cl})^{1-1/s}-(\epsilon^*/\Lambda_{\rm cl})
= [\,(s+1)/2s\,]\,(\omega_{\rm c}/\Lambda_{\rm cl}) \;.
\end{equation}
The expressions (\ref{subo3}) -- (\ref{emax1}) hold for general
$s$.

In the sub-Ohmic case $s<1$, the formula (\ref{subo3}) is valid down
to $\epsilon=0$. We have $k^+(\epsilon\to 0)\propto
\epsilon^{-(s+1)/2s}\exp{[-a_s(\Lambda_{\rm
cl}/\epsilon)^{(1-s)/s}]}$. As a result of the high density of
low-energy excitations, the barrier crossing is completely quenched
for a symmetric system at zero temperature. In the limit 
$\omega_{\rm c}\to \infty$, the position of the rate maximum is
selfconsistently determined from Eq.~(\ref{emax1}) as
\begin{align}
\epsilon^*= [\,2s/(1+s)\,]^{s/(1-s)}\,[\,2\delta_s
\Gamma(s)\,]^{1/(1-s)}\,\omega_{\rm ph} \;.
\end{align}
In the descending flank of $k^+(\epsilon)$ above $\epsilon^*$, the
multi-phonon emission processes gradually die out as 
$\epsilon$ is further increased. Then the steepest descent rate (\ref{subo3})
ceases to be valid. For $\epsilon\gg\epsilon^*$, the influence of the
environment is weak and emission of energy is determined by the one-phonon
process \cite{book}.   

The super-Ohmic case shows opposite behaviour. Here the
low bias rate is determined by  one- or few-phonon processes. With increasing 
$\epsilon$, multi-phonon processes come into operation and form the
ascending flank of $k^+(\epsilon)$. For $\epsilon \geq \omega_{\rm
ph}$, the asymptotic multi-phonon expression (\ref{subo3}) is
valid. For $\Lambda_{\rm cl}\gg\omega_{\rm c}$, the
position of the rate maximum is found from Eq.~(\ref{emax1}) as 
\begin{align}\label{emax}
\frac{\epsilon^*}{\Lambda_{\rm
cl}}=1-\frac{s+1}{2}\frac{\omega_{\rm c}}{\Lambda_{\rm cl}}+{\mathcal
O}\left(\frac{\omega_{\rm c}^2}{\Lambda_{\rm cl}^2}\right)\;, 
\end{align}
which correctly reproduces the Ohmic result \cite{Lambda}. For fixed
$\omega_{\rm c}/\Lambda_{\rm cl}$, the value of 
$\epsilon^*/\Lambda_{\rm cl}$ slopes down with increasing $s$.
Expanding the rate around its maximum $\epsilon^*$ as in
Eq.~(\ref{expand}), we obtain 
for the width $W$ and the asymmetry parameter $A$ 
\begin{align}\label{widtht0}
W^2&=\frac{s\Lambda_{\rm cl}}{\omega_{\rm c}}\left[1-\frac{s+1}{2s}
\frac{\omega_{\rm c}}{\Lambda_{\rm cl}}+{\mathcal
O}\left(\frac{\omega_{\rm c}^2}{\Lambda_{\rm cl}^2}\right)\right]\;,\\
\label{asyt0} 
A&=\frac{(s+1)\omega_{\rm c}}{3s\Lambda_{\rm cl}}
\left[1+\frac{\omega_{\rm c}}{\Lambda_{\rm cl}}+{\mathcal
O}\left(\frac{\omega_{\rm c}^2}{\Lambda_{\rm cl}^2}\right)\right]\;.
\end{align}
In the limit $s\to1$, the width (\ref{widtht0}) reduces to the Ohmic
form (\ref{widthO}), and the asymmetry parameter is  consistent with
Eq.~(\ref{asyO}) for $K\gg 1$.  
Interestingly, the width grows and the asymmetry decreases with
increasing  $s$ for fixed $\omega_{\rm c}/\Lambda_{\rm cl}$.

\section{Quantum to classical crossover for small bias}

The ET rate for a symmetric system ($\epsilon=0$) has been the subject of 
several works using different analytical approximations and numerical methods 
\cite{Siders,Bader,Wolynes,TangLin,Chandler}. 
Here we study the rate in the unbiased case for a general spectral
density employing the steepest descent method. We are particularly interested
in  the crossover from the quantum to the classical regime.
Moreover, we show how the procedure can be extended to the case of a small 
bias.  

Since the stationary point is at $z_s=-i\hbar\beta/2$
for zero bias, it is  natural to switch to the
representation  (\ref{rateX}), where the integrand is stationary at $t_s=0$. 
Expanding the integrand of the representation (\ref{Xgen})  about $t=0$, 
we find
\begin{equation}\label{Xt0}
X(t) = (\hbar\beta/4)\Lambda_1 + (\Lambda_2/\hbar\beta)\,t^2 + 
{\cal O} (t^4)\;,
\end{equation}
where the frequency scales $\Lambda_1$ and $\Lambda_2$ are given by
\begin{align}\label{Lambda1}
\Lambda_1 &= \frac{4}{\hbar\beta}\int_0^\infty \!d\omega\,
\frac{G(\omega)}{\omega^2} \,\tanh(\textstyle{\frac{1}{4}}\hbar\beta\omega)
\; ,\\ \label{Lambda2}
\Lambda_2 &= \frac{\hbar\beta}{2}\int_0^\infty \! d\omega\,
G(\omega)\,\frac{1}{\sinh(\frac{1}{2}\hbar\beta\omega)} \;.
\end{align}
For the spectral density (\ref{spectral}), we obtain the analytical
expressions  
\begin{align}\label{L1s}
\Lambda_1 &= 8B_s(k_{\rm B}T/\hbar)  + b_s(\kappa) \,
(k_{\rm B}T/\hbar)\, \left(T/T_{\rm ph}\right)^{s-1} \; ,\\[1mm]
\label{L2s}
\Lambda_2 &=  a_s(\kappa)\,(k_{\rm B}T/\hbar)\, 
\left(T/T_{\rm ph}\right)^{s-1}  \;,  \\[3mm]
\label{bs1}
b_s(\kappa)  &= 16 \delta_s\Gamma(s-1) 
\Big\{\zeta(s-1,\, 1+\kappa) -\zeta(s-1,\,{\textstyle 
\frac{1}{2}}+\kappa)\Big\} \;,\\ \label{bs3}
a_s(\kappa) &= 2 \delta_s\Gamma(s+1) \zeta(s+1,\,{\textstyle 
\frac{1}{2}}+\kappa) \; ,
\end{align}
where $T_{\rm ph} = \hbar\omega_{\rm ph} /k_{\rm B}^{}$  and
$B_s=\delta_s\Gamma(s-1)(\omega_{\rm c}/\omega_{\rm ph})^{s-1}$. There
follows from the forms (\ref{reorg}), (\ref{Lambda1}), and (\ref{Lambda2}) 
that $\Lambda_{\rm cl}\ge\Lambda_1\ge \Lambda_2$. In the
limit $\kappa\to\infty$, the 
energies $\hbar\Lambda_1$ and $\hbar\Lambda_2$
approach the classical reorganization energy,
\begin{align}\label{class}
\lim_{\kappa\to \infty} \Lambda_1 &= \Lambda_{\rm cl}\;, &
\lim_{\kappa\to \infty} \Lambda_2 &= \Lambda_{\rm cl}\;.
\end{align}
The limits (\ref{class}) are found  either from the integral
representations or by substituting the asymptotic expansion of the zeta 
function into Eqs.~(\ref{L1s}) and (\ref{L2s}).
In the quantum regime $\kappa\ll 1$, the coefficient functions take
the simpler form 
\begin{align}\label{bqm}
b_s(0)&=16 \delta_s\Gamma(s-1) [2-2^{s-1}]\zeta(s-1)\;,\\
a_s(0)&=2\delta_s\Gamma(s+1)[2^{s+1}-1]\zeta(s+1)\;.\label{aqm}
\end{align}

Substituting Eq.~(\ref{Xt0}) into Eq.~(\ref{rateX}), the rate for zero bias 
is obtained as \cite{Siders,Bader,Chandler}
\begin{align}
k^+(T,\epsilon=0)&=\frac{\Delta^2}{4}\sqrt{\frac{\pi\hbar\beta}{\Lambda_2}}
e^{-\hbar\beta\Lambda_1/4}\;.\label{rate0}
\end{align}
This formula is a quantum generalization of the classical result (\ref{marc})
for $\epsilon=0$. The quantum features are included in the
effective solvation energies $\hbar\Lambda_1$ and
$\hbar\Lambda_2$. The energy $\hbar\Lambda_1/4$ plays the role 
of a quantum activation free energy, which is below the classical value 
$E_a = \hbar\Lambda_{\rm cl}/4$ because of nuclear tunnelling.
The energy $\hbar\Lambda_2$ accounts for
quantum effects in the attempt frequency due to bath fluctuations. 
Both quantum effects lead to an enhancement of the rate over the classical
value.
The leading correction to the formula (\ref{rate0}) arising from the 
${\cal O}(t^4_{})$-term in the expansion (\ref{Xt0}) is
small when the condition
\begin{equation}\label{val}
  32\delta_s \Big(\frac{T}{T_{\rm ph}}\Big)^{s-1}\frac{[\,
\Gamma(s+1)\,\zeta(s+1,\,\frac{1}{2} +\kappa)\,]^2_{}}{\Gamma(s+3)\,
\zeta(s+3,\,\frac{1}{2} +\kappa)} \;\gg\; 1
\end{equation}
is satisfied. Physically, this condition holds in the regime of
multi-phonon processes. It is always fulfilled in the classical
limit. In the quantum regime $\kappa\ll 1$, Eq.~(\ref{val}) corresponds  
in the Ohmic limit to the damping regime $K\gg 1 $. In the super-Ohmic
case, the condition (\ref{val}) is met for temperatures well above $T_{\rm
ph}$, whereas for a sub-Ohmic bath, it is fulfilled when $T$ is 
fairly below $T_{\rm ph}$.

For finite bias, the stationary point moves away from 
$t_s=0$. However, for a small bias 
\begin{align} \label{biasval}
\epsilon\ll(1+2\kappa)\Lambda_2\;,
\end{align}
the short-time expansion (\ref{Xt0}) for $X(t)$ can still be applied.
The stationary point is located at $t_s=i\hbar\beta\epsilon/2\Lambda_2$, and
 the rate (\ref{rateX}) is found as
\begin{align}
k^+(T,\epsilon)&=\frac{\Delta^2}{4}\sqrt{\frac{\pi\hbar\beta}{\Lambda_2}}
\exp\bigg\{-\frac{\hbar\beta\Lambda_1}{4}+\frac{\hbar\beta\epsilon}{2}
-\frac{\hbar\beta\epsilon^2}{4\Lambda_2}\bigg\} 
\;.\label{ratee}
\end{align}
This expression describes the
crossover from the classical to the quantum regime for general spectral
densities. Corrections to the formula (\ref{ratee})
are small if the conditions (\ref{val}) and (\ref{biasval}) are
fulfilled.
In the classical regime $\kappa\gg 1$,
the condition (\ref{biasval}) is satisfied even when the bias is very
large. We see from Eq.~(\ref{class}) that the expression (\ref{ratee})
reduces to the classical rate (\ref{marc}) as $\kappa\to\infty$.
As the temperature is lowered to the quantum regime, the particular 
properties of the spectral density $G(\omega)$ become 
relevant which indicates that nuclear tunnelling becomes effective.
The quantum effects are captured by the deviations of the effective 
solvation energies $\hbar\Lambda_1$ and $\hbar\Lambda_2$ from 
$\hbar\Lambda_{\rm cl}$. 
However, the bias range (\ref{biasval}) in which the rate formula
(\ref{ratee}) is valid shrinks as temperature is lowered. In the quantum
regime $\kappa\ll 1$, the condition (\ref{biasval}) reduces to 
$\epsilon\ll \Lambda_2\propto T^s$. Hence the bias range goes to zero as
$T\to 0$. In the quantum regime $\kappa\ll 1$, the formula (\ref{ratee}) 
does not properly describe the rate around the maximum.  

In the Ohmic case $s\to 1$, we obtain from Eqs.~(\ref{L1s}) and (\ref{L2s})
in the quantum regime $\kappa\ll 1$
\begin{align}
\Lambda_1&=(8K/\hbar\beta)\ln(\hbar\beta\omega_{\rm c}/\pi)\;, &
\Lambda_2&=\pi^2 K/\hbar\beta\;,
\end{align}
and the rate expression (\ref{ratee}) takes the form
\begin{align}\label{sdro}
  k^+(T,\epsilon)&=\frac{\Delta^2}{4\omega_{\rm c}}\sqrt{\frac{\pi}{K}}
         \left(\frac{\pi}{\hbar\beta\omega_{\rm c}}\right)^{2K-1}
\exp\bigg\{\frac{\hbar\beta\epsilon}{2}-\frac{1}{K}
   \left(\frac{\hbar\beta\epsilon}{2\pi}\right)^2\bigg\}\;.   
\end{align}
Now, the conditions (\ref{val}) and (\ref{biasval}) reduce to
$K\gg 1$ and $\hbar\epsilon\ll\pi^2K\, k_{\rm B}T$. 
In this parameter regime, the steepest descent-rate (\ref{sdro}) coincides
with the previous expression (\ref{ROT}). 

For a larger bias than given by Eq.~(\ref{biasval}), the
stationary point is no longer near to 
$t=0$ and the expansion (\ref{Xt0}) cannot be employed any more. This
case is studied in the next section.

\section{Quantum to classical crossover for strong bias}

As the bias is further increased, the stationary point $z_s$ approaches the
origin of the complex time plain, and for $\epsilon=\Lambda_{\rm cl}$,
we have $z_s=0$.  
In general, for large bias and/or high temperature, the stationary point
is near $z=0$. To explore this regime in which quantum signatures are 
most pronounced, we expand the integral representation (\ref{Qgen}) of 
$Q(z)$ around $z=0$ up to fourth order in $z$,
\begin{equation}\label{Q4}
Q(z) = i\Lambda_{\rm cl}z +\frac{c_2(\kappa)}{2!}\,(\omega_{\rm c} z)^2
-i\,\frac{c_3}{3!}\,(\omega_{\rm c} z)^3 - \frac{c_4(\kappa)}{4!}\,
(\omega_{\rm c} z)^4 + {\cal O}(\omega_{\rm c} z)^5\;.
\end{equation} 
The dimensionless coefficient functions $c_j(\kappa)$ are given by
\begin{align}
c_2(\kappa) &=\frac{1}{\omega_{\rm c}^2}\int_0^\infty\!
d\omega\,G(\omega)\coth({\textstyle\frac{1}{2}}\hbar\beta\omega) \;,\notag\\
c_3 &=\frac{1}{\omega_{\rm c}^3}\int_0^\infty\! d\omega\,\omega
G(\omega)\;,\label{cgen}\\
c_4(\kappa) &=\frac{1}{\omega_{\rm c}^4}\int_0^\infty\!
d\omega\,\omega^2
G(\omega)\coth({\textstyle\frac{1}{2}}\hbar\beta\omega) \;.\notag
\end{align}
For the spectral density 
(\ref{spectral}), we readily obtain 
\begin{align}\nonumber
c_2(\kappa) &= s\,[\,1 +2 \kappa^{s+1}_{}\zeta(s+1,1+\kappa)\,]\,
(\Lambda_{\rm cl}/\omega_{\rm c}) \;, \\ \label{c}
c_3 &= s(s+1)(\Lambda_{\rm cl}/\omega_{\rm c}) \;,\\
c_4(\kappa) &= s(s+1)(s+2)\,[\,1+ 2\kappa^{s+3}_{}
\zeta(s+3,1+\kappa)\,]\,
(\Lambda_{\rm cl}/\omega_{\rm c})\;.\nonumber
\end{align}

In earlier works, the analysis of the rate has been based on truncating the
series (\ref{Q4}) at order $z^{2}$ \cite{Siders,Garg}. However, this
crude approximation, which has been often termed ``semiclassical'', 
fails to describe the 
actual shift of the rate maximum $\epsilon^{*}$ away from $\Lambda_{{\rm cl}}$
and  the asymmetric distribution around the maximum in the quantum regime
$\kappa\ll 1$, as noticed already in Ref.~\cite{Ando}.
In the sequel, we study the quantum effects of the rate related to 
terms of order $z^{3}$ and higher in Eq.~(\ref{Q4}).

In the bias regime
\begin{equation}
  \label{biasdev}
  |\Lambda_{\rm cl}-\epsilon|\ll\omega_{\rm c}\, c_2^2(\kappa)/c_3(\kappa)\;,
\end{equation}
the stationary point is obtained from  Eq.~(\ref{transc}) with
Eq.~(\ref{Q4}) in the form of a series in $\Lambda_{\rm
cl}-\epsilon$. Up to second order, we find   
\begin{equation}\label{station}
z_s=-\frac{i}{\omega_{\rm c}}\left(\frac{\Lambda_{\rm
cl}-\epsilon}{c_2\omega_{\rm c}}\right) 
-\frac{i}{2\omega_{\rm c}}\frac{c_3}{c_2}\left(\frac{\Lambda_{\rm
cl}-\epsilon}{c_2\omega_{\rm c}}\right)^2 \;.
\end{equation}
Then the rate is found in the form (\ref{SDR}), 
\begin{align}
k^{+}(T, \epsilon) &= \frac{\Delta^2}{4} 
{\left( \frac{2\pi}{ F''(z_s)} \right )}^{1/2}
\exp[- F(z_s)] \; ,\notag\\
F(z_s)&=\frac{1}{2c_2}\left(\frac{\Lambda_{\rm cl}-\epsilon}{\omega_{\rm c}}
\right)^2 +\frac{c_3}{6c_2^3}\left(\frac{\Lambda_{\rm
cl}-\epsilon}{\omega_{\rm c}}\right)^3 \;,\label{Fz}  \\ 
F''(z_s)&=\left[c_2-\frac{c_3}{c_2}\left(\frac{\Lambda_{\rm
cl}-\epsilon}{\omega_{\rm c}}\right)-\frac{1}{2}
\left(\frac{c_3^2}{c_2^3}-\frac{c_4}{c_2^2}\right) 
\left(\frac{\Lambda_{\rm
cl}-\epsilon}{\omega_{\rm c}}\right)^2\right]\omega_{\rm c}^2
\;.\notag
\end{align}
This rate expression together with Eq.~(\ref{cgen}) describes the full
crossover from quantum-mechanical to classical behaviour for a general 
spectral density. It is valid
if the bias condition (\ref{biasdev}) and the steepest descent
condition (\ref{validity}), which  approximately takes the form
$c_4/8c_2^2\ll1$, are fulfilled. For $\kappa\gg 1$, we see from
Eq.~(\ref{cgen}) that these conditions are fulfilled for any bias and
spectral density, and the
rate reduces to the classical result (\ref{marc}). For lower
$\kappa$, the rate expression (\ref{Fz}) shows substantial
deviations from the classical behaviour. 
The quantum activation factor $\exp[-F(z_s)]$ of
the rate alone would  predict activationless transfer for
$\epsilon=\Lambda_{\rm cl}$. However, the rate maximum is actually
shifted by the fluctuation factor $F''(z_s)$. We consistently obtain 
\begin{align}\label{eps*t4}
  \frac{\epsilon^*}{\Lambda_{\rm cl}}&=
   1-\frac{c_3}{2c_2}\frac{\omega_{\rm c}}{\Lambda_{\rm cl}} \;.
\end{align}
The rate around the maximum is again described by the form (\ref{expand}).
The width $W$ and the asymmetry parameter $A$ are given by 
\begin{align}
   W^2&=c_2\left[1+\frac{1}{2}\left(\frac{c_3^2}{c_2^3}
      -\frac{c_4}{c_2^2}\right)\right] \label{wt4}\;,\\
\label{asyt4}
 A&=\frac{c_3}{3c_2^2}\;.
\end{align}
We emphasize that the expressions (\ref{eps*t4}) -- (\ref{asyt4})
together with Eq.~(\ref{cgen}) are valid for a general spectral density. 
In the following, we will restrict ourselves to the special form
(\ref{c}) of the coefficient functions $c_j$. Before discussing the
quantum signatures in the parameters $\epsilon^*$, $W$, and $A$ in
detail, we now consider the conditions (\ref{biasdev}) and
(\ref{validity}) in the quantum regime $\kappa\ll 1$.
First, the steepest descent condition
(\ref{validity}) is approximately given by
$\Lambda_{\rm cl}\gg\omega_{\rm c}$. This condition is practically
always satisfied in the super-Ohmic case when 
$\omega_{\rm c}\gg\omega_{\rm ph}$, whereas in the Ohmic case, it
corresponds to the regime $K\gg 1$. It is usually not fulfilled in the
sub-Ohmic case, except for very strong coupling $2\delta_s\Gamma(s)\gg
(\omega_{\rm c}/\omega_{\rm ph})^{1-s}$. 
Secondly, the bias condition (\ref{biasdev}) takes roughly the form 
$|\Lambda_{\rm cl}-\epsilon| \ll \Lambda_{\rm cl}$, and thus at $T=0$ we are
confined to a narrow bias interval around $\Lambda_{\rm cl}$ \cite{kappa}. 
Importantly, the most interesting area around the rate maximum
falls into this interval. It is easily verified that in this bias interval
the $T=0$ limit of the rate (\ref{Fz}) reduces to the previous result
(\ref{subo3}). 

\begin{figure}[h]
\setlength{\unitlength}{1cm}
\centerline{
\begin{picture}(6.5,6.5)
  \put(0,0.5){\epsfig{file=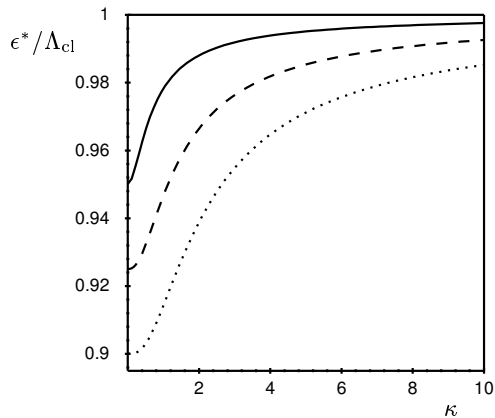,width=6.5cm,angle=0}}
\end{picture}}
\caption{\label{epsmax}$\epsilon^*/\Lambda_{\rm cl}$ as a function of 
$\kappa$ for $\omega_{\rm c}/\Lambda_{\rm cl}=0.05$. The solid, dashed, 
and dotted lines correspond to the cases $s=1$, 2, and 3,
respectively.} 
\end{figure}
The position of the rate maximum,  Eq.~(\ref{eps*t4}),
is plotted as a function of $\kappa$  for different values of $s$ in
Fig.~\ref{epsmax}. 
In the asymptotic low temperature regime, we find from Eq.~(\ref{eps*t4}) 
\begin{align}\label{eps*t4low}
\frac{\epsilon^*}{\Lambda_{\rm
cl}}&=1-\frac{s+1}{2}\frac{\omega_{\rm c}}{\Lambda_{\rm cl}} 
      \left[1-2\zeta(s+1)\bigg(\frac{k_{\rm B}T}{\hbar\omega_{\rm
      c}}\bigg)^{s+1}\right]\;,    
\end{align}
which correctly reproduces the $T=0$ result, Eq.~(\ref{emax}). The leading 
low-temperature corrections vary as $T^{s+1}$. For $s=1$, the form
(\ref{eps*t4low}) reduces to the earlier result (\ref{eps*O}) for
$K\gg1$. 
With increasing $\kappa$, the maximum $\epsilon^{*}$ is shifted upwards
until it finally reaches the classical value $\Lambda_{\rm
  cl}$. With increasing  $s$, the approach to the classical limit is
more gradual. As visible in Fig.~\ref{epsmax}, the deviations
from the classical behaviour are more distinct for larger $s$ throughout
the whole temperature range.

\begin{figure}[h]
\setlength{\unitlength}{1cm}
\centerline{
\begin{picture}(6.5,6.5)
  \put(0,0.5){\epsfig{file=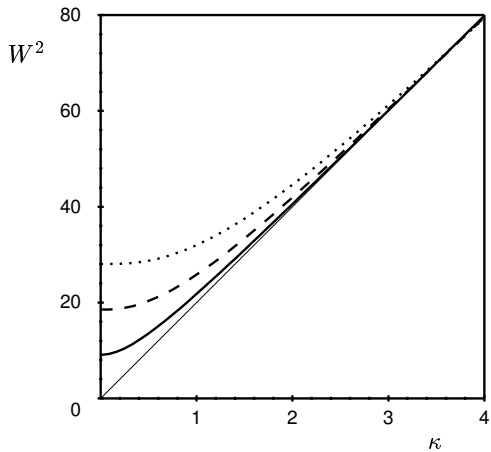,width=6.5cm,angle=0}}
\end{picture}}
\caption{\label{wi}The squared width $W^2$
   as a function of $\kappa$ for $\omega_{\rm c}/\Lambda_{\rm cl}=0.1$.
 The solid, dashed, and dotted lines correspond to the cases $s=1$, 2,
 and 3, respectively.  The thin solid line represents the classical component 
$W_{\rm cl}^{2}=2(\Lambda_{\rm cl}/\omega_{\rm c})\kappa$. The
difference between this straight line and the full 
squared width is the quantum component $W_{\rm qm}^{2}$.}
\end{figure}

\begin{figure}[h]
\setlength{\unitlength}{1cm}
\centerline{
\begin{picture}(6.5,6.5)
  \put(0,0.5){\epsfig{file=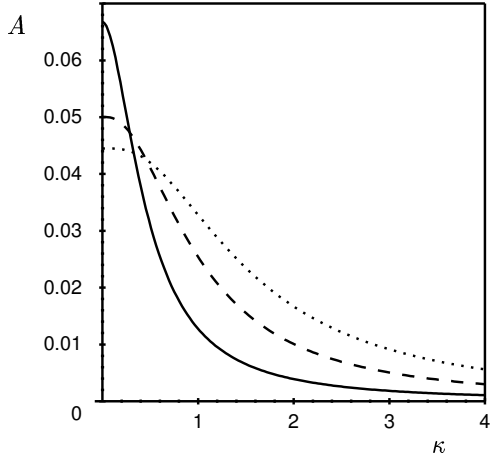,width=6.5cm,angle=0}}
\end{picture}}
\caption{\label{asy}The asymmetry parameter $A$  
as a function of $\kappa$ for $\omega_{\rm c}/\Lambda_{\rm cl}=0.1$.
 The solid, dashed, and dotted lines correspond to the cases  $s=1$, 2,
 and 3, respectively. Since $A>0$, the  wing of the rate above the
maximum is enhanced compared to the wing below the maximum. In the quantum 
regime $\kappa\ll 1$, the deviations from the
classical value $A=0$ are more distinct for smaller $s$. This is in
contrast to the behaviour of $\epsilon^*$ and $W^2$ for $\kappa\ll 1$
displayed in Figs.~\ref{epsmax} and \ref{wi}. }
\end{figure}

The width and the asymmetry parameter of the rate 
are displayed in Figs.~\ref{wi} and \ref{asy}.

At very low temperatures, the expressions (\ref{wt4}) and
(\ref{asyt4}) take the form
\begin{align}\label{widthlow}
  W^2&=\frac{s\Lambda_{\rm cl}}{\omega_{\rm c}}\bigg(1-\frac{s+1}{2s}
      \frac{\omega_{\rm c}}{\Lambda_{\rm cl}}\bigg)
   \left[1+2\zeta(s+1)\bigg(1-\frac{s^2-1}{2s}\frac{\omega_{\rm c}}
   {\Lambda_{\rm cl}}\bigg)\bigg(\frac{k_{\rm B}T}{\hbar\omega_{\rm
      c}}\bigg)^{s+1}\right]\;,\\
  \label{asylow} A&=\frac{(s+1)\omega_{\rm c}}{3s\Lambda_{\rm cl}}
      \left[1-4\zeta(s+1)\bigg(\frac{k_{\rm B}T}{\hbar\omega_{\rm
      c}}\bigg)^{s+1}\right] \,.
\end{align}
Again, the $T=0$ results (\ref{widtht0}) and
(\ref{asyt0})  are correctly reproduced \cite{noteA}. With increasing
$s$, the width grows, whereas the asymmetry decreases.   
The low temperature corrections  vary as $T^{s+1}$. For $s\to 1$,
we recover the Ohmic results (\ref{widthO}) and (\ref{asyO}) for $K\gg 1$.
With increasing
temperature, the asymmetry decreases and the width grows. However, the 
increase of the width is due to the ``classical component'' $W_{\rm
cl}^2=2(\Lambda_{\rm cl}/\omega_{\rm c})\kappa$ arising from thermal
activation, whereas the ``quantum component'' $W^2_{\rm qm}$ decreases.
For this reason, the quantum effects are reduced with increasing temperature.
In the limit $\kappa\to \infty$, the asymmetry finally drops 
down to zero, and the width approaches the classical form
$W_{\rm cl}^{2}=2(\Lambda_{\rm cl}/\omega_{\rm c})\kappa$.
With increasing $s$, the curves in Figs.~\ref{wi} and \ref{asy}
approach the classical limits more gradually, and the deviations from
the classical behaviour are more pronounced. This is in  
correspondence with the behaviour of $\epsilon^{*}$. Thus, with growing $s$,
the quantum regime extends to higher temperatures. Clearly, this is
due to the larger contribution of high frequency modes leading to
an enhancement of nuclear tunnelling. 

\section{Conclusions}

We have studied the nonadiabatic tunnelling rate in the biased spin-boson 
model within the Golden Rule formula. We have analyzed the
quantum signatures of the rate in the position of the maximum $\epsilon^{*}$,
the width $W$, and the asymmetry parameter $A$ for
different spectral densities. In the Ohmic case, we have calculated the rate
in analytic form at $T=0$, and we have presented the
leading low temperature corrections. Since the rate maximum is located in 
the strong bias regime, it has been essential to treat the bath
correlation function  beyond the scaling limit.
For an Ohmic and non-Ohmic spectral density, we have investigated the entire
quantum to classical crossover of the rate within steepest descent,
and we have discussed the parameter bounds within which this approximation is
justified. We wish to emphasize that the characteristic quantum features 
found are not artefacts of the steepest descent method.
For the parameter regions displayed in Figs.~\ref{epsmax} -- \ref{asy}, the
deviations of the analytical expressions from numerical computations of 
the Golden Rule rate 
are negligibly small. 
The quantum signatures  of the rate are gradually extincted with
increasing temperature. We found that for fixed low temperature,
$\kappa\ll 1$, the
quantum features in the position of the maximum and the width are growing with
increasing $s$. In contrast, the quantum signatures in the asymmetry parameter
become less pronounced if we switch from Ohmic to super-Ohmic.
At higher temperatures, $\kappa>1$, the quantum signatures in the parameters 
$\epsilon^*$, $W$, and $A$ are more distinct for larger $s$. Thus in
the super-Ohmic case, the regime in which deviations from the
classical behaviour are significant extends to higher temperatures than 
in the Ohmic or sub-Ohmic case. 

Many of our results are not limited to the particular
form (\ref{spectral}) of the spectral density. They also apply to spectral 
densities with a peak-like structure at higher frequencies, which are relevant
when inner-sphere reorganization or specific vibrational modes in the 
solvent are important. The quantum effects studied here are particularly 
pronounced for solvent modes  with energy above the thermal energy. 
Finally, our study is also relevant for single charge transfer in 
ultrasmall junctions in which the electromagnetic environment plays the role 
of the solvent \cite{ingold,book}.
    
\acknowledgments
E.~P. thanks G. Giaquinta for constant encouragement, and  
acknowledges financial support by the INFM under the PRA-QTMD programme
and by the Sonderforschungsbereich 382 of the Deutsche 
Forschungsgemeinschaft (Bonn).


\begin{references}
\bibitem{marcus1} For a review see R.A. Marcus and N. Sutin, 
        Biochim. Biophys. Acta {811} (1985) 265, and references
        therein.

\bibitem{Ulstrup} J. Ulstrup, {\it Charge Transfer Processes in Condensed
Media}
        (Springer, Berlin, 1979).

\bibitem{Fain} B. Fain, {\it Theory of Rate Processes in Condensed Media}
         (Springer, Berlin, 1980).

\bibitem{Benderskii} A. Benderskii, D.E. Makarov, and C.A. Wright,
Adv. Chem. Phys. 88 (1994) 1.

\bibitem{Devault} D. Devault, Q. Rev. Biophys. Acta {977} (1989) 99.

\bibitem{marcus2} R.A. Marcus, J. Chem. Phys. {24} (1956) 966.

\bibitem{levich} V.G. Levich, in {\em Advances in Electrochemistry
and Electrochemical Engineering}, ed. by P. Delahay and C.W. Tobias,
Vol. 4, p.249 (Interscience, New York, 1966).


\bibitem{hanggi} P. H\"anggi, P. Talkner, and M. Borkovec, Rev. Mod. Phys.
        {62} (1990) 251. 

\bibitem{Siders} P. Siders and R.A. Marcus, J. Am. Phys. Soc. {103}
(1981) 741. 

\bibitem{Bader} R.A. Kuharski, J.S. Bader, D. Chandler, M. Sprik, 
        M.L. Klein, and R.W. Impey, J. Chem. Phys. {89} (1988) 3248;
        J.S. Bader, R.A. Kuharski, and D. Chandler, J. Chem. Phys.
        {93} (1990)~1. 

\bibitem{Wolynes} P.G. Wolynes, J. Chem. Phys. {87} (1987) 6559.


\bibitem{Garg} A. Garg, J.N. Onuchic, and V. Ambegaokar, J. Chem. Phys.
        {83} (1985)  4491. 

\bibitem{TangLin} J. Tang and S.H. Lin, J. Chem. Phys. {107} 
        (1997) 3485.

\bibitem{Chandler} D. Chandler, in 
        {\it Liquids, Freezing, and the Glass Transition}, Les Houches 
        lectures,
        edited by  D. Levesque, J.P. Hansen, and J. Zinn-Justin (Elsevier
        Science, North Holland, 1991).

\bibitem{leggett} A.J. Leggett, S. Chakravarty, A.T. Dorsey, M.P.A. Fisher,
           A. Garg, and W. Zwerger, Rev. Mod. Phys. {59} (1987) 1;
          {\it ibid.} {67} (1995) 725  [erratum].

\bibitem{book} U. Weiss, {\it Quantum Dissipative Systems},
           Series in Modern Condensed Matter Physics, Vol.2,
           (World Scientific, Singapore, second edition, 1999).

\bibitem{grabert} H. Grabert, Phys. Rev. B {46} (1992) 12\,753.

\bibitem{Grad} L.S. Gradshteyn, L.M. Ryzhik,
        {\it Tables of Integrals, Series and Products}, (Academic Press,
        New York, 1980).  

\bibitem{ohmic1} U. Weiss and H. Grabert, Phys. Lett. {108 A} (1985) 63.

\bibitem{ohmic3}
        H. Grabert and U. Weiss, Phys. Rev. Lett. {54} (1985) 1605; 
     M.P.A. Fisher and A.T. Dorsey, Phys. Rev. Lett. {54} (1985) 1609.
        

\bibitem{ohmic2} C. Aslangul, N. Poitier, and D. Saint-James, J. Phys.
(Paris) {47} (1986) 1671. 


\bibitem{EMW} R. Egger, C.H. Mak, and U. Weiss, J. Chem. Phys. {100} 
              (1994) 2651. 

\bibitem{Lambda} Eq.~(\ref{emax}) is also valid in the
sub-Ohmic case if the condition $\Lambda_{\rm cl}\gg\omega_{\rm c}$
is fulfilled. However, for $s<1$ this condition is usually 
only satisfied for very strong coupling (cf. Section VII).

\bibitem{Ando} K. Ando, J. Chem. Phys. {106} (1997) 116.
 
\bibitem{kappa} With increasing temperature, the $T=0$ conditions
$\Lambda_{\rm cl} \gg \omega_{\rm c}$ and $|\Lambda_{\rm cl}-\epsilon| \ll
\Lambda_{\rm cl}$ are softened.

\bibitem{noteA} Actually, the form (\ref{asylow}) at $T=0$ gives
Eq.~(\ref{asyt0}) only in lowest order of $\omega_{\rm c}/\Lambda_{\rm
cl}$. To reproduce the second order term in 
Eq.~(\ref{asyt0}), we have to  expand $Q(z)$ up to 5th order in $z$. 

\bibitem{ingold} G.-L. Ingold and Yu.V. Nazarov, in 
{\em Single Charge Tunneling},
ed. by H. Grabert and M.H.~Devoret, NATO ASI Series B: Physics Vol.~294
(Plenum Press, New York, 1992). 

\end{references}
\end{document}